\documentclass{JHEP3}
\usepackage{amssymb,amsfonts,amsmath,amsthm,latexsym,graphicx,verbatim,epsfig}

\newcommand{\nn}{\nonumber}
\newcommand{\beq}{\begin{equation}}
\newcommand{\eeq}{\end{equation}}
\newcommand{\bea}{\begin{eqnarray}}
\newcommand{\eea}{\end{eqnarray}}

\newcommand{\1}{ \,  \raisebox{+0.14em}{{\hbox{{\rm \scriptsize ]}} \raisebox{-0.2em}{\kern-.8em\hbox{1}}}} \, }  

\title{\center{Equidistribution Rates,  Closed String Amplitudes, and the Riemann Hypothesis.}}

\author{Sergio L. Cacciatori$^{a}$ and
Matteo Cardella$^{b}$ \\
$^a$ Department of Physics and Mathematics, \\
\hspace*{0.15cm} Universit\`a dell'Insubria, Via Valleggio 11, I-22100 Como, Italy, and \\
\hspace*{0.15cm} INFN, Sezione di Milano, Via Celoria 16, I-20133 Milano, Italy \\
\hspace*{0.15cm} E-mail address: sergio.cacciatori@uninsubria.it.

\

$^b$ Department of Physics, Universit\`a di Milano Bicocca, \\
\hspace*{0.15cm} piazza della Scienza 3, I-20126 Milano, Italy, and\\
\hspace*{0.15cm} Erwin Schr\"odinger International Institute for Mathematical Physics, \\
\hspace*{0.15cm} Boltzmanngasse 9 A-1090, Vienna, Austria \\
\hspace*{0.15cm} E-mail address: matteo@phys.huji.ac.il.
}

\abstract{We study asymptotic relations connecting  unipotent averages of $Sp(2g,\mathbb{Z})$
automorphic forms to their integrals over the moduli space of  principally polarized abelian varieties.
 We obtain  reformulations of the Riemann hypothesis  as a class of problems concerning the
computation of  the equidistribution convergence rate
 in those asymptotic relations. We discuss applications of our results to
 closed string amplitudes. Remarkably,  the Riemann hypothesis can
 be rephrased in terms of ultraviolet  relations occurring in  perturbative
 closed string theory.}


\begin{document}

\section{Introduction}
 It is well known  that  long horocycles tend to become uniformly distributed in the unit tangent bundle of the  modular surface $Sl(2, \mathbb{Z})  \backslash Sl(2,\mathbb{R})$, \cite{F},\cite{Dani}. For a large class of homogenous spaces given by quotients of Lie groups by  discrete subgroups,
 Ratner  theorems on
  equidistribution of unipotent flows \cite{Ratner1},\cite{Ratner2} provide the key  for proving  striking results in number theory.
  Thanks to equidistribution, the  unipotent average of an automorphic function $f$ in a suitable limit
  converges to its average on the homogenous space.
   A very interesting quantity is the $f$ convergence rate for its connection to the (Grand)Riemann hypothesis.
   This quantity is not provided by Ratner algebraic   methods.
   For  the $Sl(2, \mathbb{Z})  \backslash Sl(2,\mathbb{R})$ modular surface,  Zagier \cite{ZagierE}, (see also \cite{Zagier},\cite{SarnakH},
   \cite{Verjovsky} as related works) has shown that  if  just in a single case of  $Sl(2,\mathbb{Z})$-invariant smooth function $f$
 of rapid decay as $z \rightarrow i\infty$   the horocycle  convergence rate is  $O(\Im(z)^{3/4})$ as $\Im(z) \rightarrow 0$  then the Riemann hypothesis is true!
 Information contained in the modular surface $Sl(2, \mathbb{Z})  \backslash Sl(2,\mathbb{R})$       on the Riemann zeta function $\zeta(s)$
are due to the arithmetic nature of this surface, being the space of inequivalent  unimodular latices on the plane $\mathbb{R}^2$.

In this paper we consider  $Sp(2g, \mathbb{Z})$ automorphic forms $f$  defined on the genus $g$ Siegel upper space $\mathbb{H}_g $.
 By extending  old ideas of  Rankin \cite{Rankin} and Selberg \cite{Selberg}, we study by analytic methods  equidistribution of the $f$ average along unipotent flows. A great simplification follows by the use of Iwasawa decomposition for $Sp(2g, \mathbb{R})$
 in the unfolding of the modular domain $Sp(2g,\mathbb{Z})\backslash Sp(2g, \mathbb{R})$. By using analytic properties of
 genus $g$  Eisenstein series associated to various components of the boundary of the modular domain,
 we obtain that  certain values of the convergence rates of the unipotent averages are compatible only with the
Riemann hypothesis. Indeed, our results provide   reformulations of the Riemann hypothesis in terms
of a class of problems for the  convergence rate in  the  unipotent dynamics in the $Sp(2g,\mathbb{Z}) \backslash \mathbb{H}_g$ modular space.
These results  generalize to modular domains $Sp(2g,\mathbb{Z}) \backslash  Sp(2g,\mathbb{R})$ the
  $O(\Im(z)^{3/4})$ result    \cite{ZagierE} for the genus $g=1$ modular surface  $Sl(2,\mathbb{Z}) \backslash Sl(2,\mathbb{R})$.

Among the motivations for our work, there is a certain relevance of equidistribution properties of unipotent flows in
String Theory \cite{Cardella:2008nz}. Indeed, uniform distribution is behind some deep UV/IR properties of perturbative   String Theory.
Equidistribution of large horocycles in $Sl(2,\mathbb{Z}) \backslash Sl(2,\mathbb{R})$
 is at the heart \cite{Cardella:2008nz} of the  UV/IR connection   among asymptotic Fermi-Bose degeneracy and space-time stability \cite{Kutasov:1990sv}.
In a different direction, equidistribution of large horocycles involving congruence subgroups $\Gamma_{0}[N] \subset Sl(2,\mathbb{Z})$ modular domains
allows to write   a $\mathbb{Z}_N$ orbifold  torus amplitude  as a certain limit of a one-dimensional integral \cite{Cardella:2008nz}.
This may lead to some simplifications for computing one loop torus integrals for $\mathbb{Z}_N$ orbifolds.

We find very appealing the fact that in some known example the equidistribution convergence rate
corresponds to some quantity in  String Theory.
As an example, in \cite{ACER} the genus $g =1$ case  is studied  for non tachyonic closed string backgrounds
admitting a CFT description. In those cases  the supertrace over the closed string states is a  $Sl(2,\mathbb{Z})$ automorphic function,
and the horocycle average of the automorphic supertrace counts the difference among the total number of closed strings
bosonic minus fermionic excitations below an ultraviolet cutoff controlled by the horocycle radius.

Indeed, a very interesting direction which  motivates our work
is  the search of examples in String theory where, besides the correspondence   of the
equidistribution rate to some ``physical'' quantity, also dualities which allow to
 map the rate are available. Then one  may attempt to leave modularity
by using a string duality and attempt to estimate the equidistribution rate from the
other side of the duality. This latest possibility  seems quite appealing and it is yet unexplored.

In the last part of this paper we briefly outline various possible  applications to String Theory of our $Sp(2g, \mathbb{R})$ results.
Some of those applications  are presently under investigation and will be presented  in future publications.
 Our results allow to re-express genus $g$ closed string amplitudes  as lower dimensional integral along
unipotent flows. This has  interesting applications for $g = 2$ closed string amplitudes \cite{DP0},\cite{DP1},\cite{DP2},\cite{DP3},\cite{DP4},\cite{DP5}, and in recently  proposed expression for genus $g \ge 3$
closed string amplitudes \cite{Cacciatori:2007vk},\cite{Cacciatori:2008ay},\cite{Cacciatori:2008pj},\cite{Piazza:2008qv},\cite{Grushevsky:2008zm} \cite{Morozov:2008wz}.
 The main idea is to use equidistribution results in order to obtain constraints deriving from
 finiteness of genus $g$ closed string amplitudes.
 This would give a  genus  $g$ generalizations of  the genus one  asymptotic supersymmetry
 constraint for non tachyonic closed string spectra originally obtained in  \cite{Kutasov:1990sv}.

\section{Equidistribution theorems}

The genus $g$  Siegel upper space  $\mathbb{H}_{g} \subset Mat(g, \mathbb{C})$ is the set  of
complex $g \times g$ symmetric  matrices with  positive definite imaginary part $\mathbb{H}_{g} = \{\tau \in Mat(g, \mathbb{C}) | \tau = \tau^{t},  \Im(\tau) > 0    \}$.
$\mathbb{H}_{g}$ is isomorphic to the Lie  coset $ \mathbb{H}_g \simeq Sp(2g,\mathbb{R}) / \left( Sp(2g,\mathbb{R}) \cap SO(2g, \mathbb{R}) \right)$.
For $m$ in the coset
\beq
m =
 \begin{pmatrix}
  a   &    b \\
  c   &     d  \\
\end{pmatrix},
\eeq
the bijective map is given by
\beq
\tau(m) = (ai\mathbb{I} + b)(ci\mathbb{I} + d)^{-1}. \label{map}
\eeq
The Iwasawa decomposition
allows to write a symplectic  matrix $g$ in $Sp(2g,\mathbb{R})$   as  $g = nak$,  $k \in  SO(2g, \mathbb{R}) \cap Sp(2g, \mathbb{R})$, $a$ is a positive definite diagonal  matrix
and $n$ is a unipotent matrix.
It is convenient to employ the following $g \times g$ blocks parametrization
\beq
a =  \begin{pmatrix}
 V & 0 \\
  0   &   V^{-1}
\end{pmatrix}, \qquad   V =  diag\left(\sqrt{v_{1}}, \dots , \sqrt{v_{g}}\right),
\eeq
for the Abelian part with $v_{i} > 0$, $i = 1,\dots, g$, and
\beq
n =
\begin{pmatrix}
U & \, \, WU^{-t} \\
0   & \, \, U^{-t}
\end{pmatrix}
\eeq
for the unipotent part, with $W$ symmetric real $g \times g$ matrix
\beq
W =
\begin{pmatrix}
w_{11}  & w_{12} & \dots  &  w_{1g} \\
 w_{12} & w_{22} & \dots  & w_{2g}  \\
\vdots  & \vdots & \ddots & \vdots  \\
w_{1g}  &  w_{2g}& \dots  &  w_{gg} \\
\end{pmatrix}  \nn 
\eeq
and $U$ upper triangular  real $g \times g$ matrix

\beq
U =
\begin{pmatrix}
 1    &   u_{12}        & \dots  &   u_{1g}   \\
 0    &   1 & \dots  &      u_{2g}  \\
\vdots  & \vdots & \ddots & \vdots  \\
 0    &  0  & \dots  &  1 \\
\end{pmatrix}.  \nn   
\eeq
With the above parametrization, $\mathbb{H}_{g}$ in Iwasawa coordinates is given by  (\ref{map})
\beq
\tau(m) = W + iUV^{2}U^{t}. \label{Iwcoord}
\eeq

\

We are interested in  $\Gamma_{g} \sim Sp(2g,\mathbb{Z})$ automorphic forms,
and in particular in proving equidistribution of
subgroup flows in order to reduce a typical    modular integral
\beq
\int_{\Gamma_{g} \backslash \mathbb{H}_{g}} d\mu f(\tau),
\eeq
into a lower dimensional integral.
 $f$ is a weight zero  automorphic function under the modular group $\Gamma_g \sim Sp(2g,\mathbb{Z})$,
   $f(\gamma(\tau)) = f(\tau)$, $\gamma \in \Gamma_g$, and $d\mu$
is the $\Gamma_{g}$-invariant hyperbolic $\mathbb{H}_{g}$ measure
\beq
d\mu = \frac{1}{det(\Im(\tau))^{g+1}}\prod_{i \le j} d \, \Re(\tau)_{ij} d \, \Im(\tau)_{ij},
\eeq
where $\tau_{ij} = \Re(\tau)_{ij} + i \Im(\tau)_{ij}$.

\vspace{.5 cm}

The  general strategy will be to consider the following modular integral
\beq
I_{g,r}(s) = \int_{\Gamma_{g} \backslash \mathbb{H}_{g}} d\mu f(\tau) \phi_{r}(\tau,s), \label{modularI}
\eeq
where $\phi_{r}(\tau,s)$ is a suitable  genus $g$ Poincar\'e series.
An interesting family of Poincar\'e series has the form

\beq
\phi_{r}(\tau,s) = \sum_{\Gamma_{g} \cap P_{g - r}  \backslash \Gamma_{g}}\varphi \left( \frac{det(\Im(\gamma(\tau)))}{det(\Im((\gamma (\tau)_{22}))}, s  \right), \label{Poinc}
\eeq
where we use the following block decomposition
\beq
\tau =
\begin{pmatrix}
\tau_{11} & \tau_{12} \\
\tau_{12}^{t}    & \tau_{22}
\end{pmatrix},
\eeq
with $\tau_{11} \in \mathbb{H}_{r}$ and $\tau_{22} \in \mathbb{H}_{g - r}$, $1 \le r  \le g - 1$.
$P_{g - r} \subset Sp(2g,\mathbb{R})$ is the  parabolic subgroup which stabilizes the $(g - r)$-dimensional (in the complex sense)
rational component of the boundary of $\Gamma_{g}  \backslash \mathbb{H}_{g}$, (see the appendix for an account on the properties
of the $\Gamma_{g}  \backslash \mathbb{H}_{g}$ boundary and related parabolic subgroups of $Sp(2,\mathbb{R})$).
\\
From (\ref{Iwcoord}) it follows that
\beq
det(\Im(\tau)) = \prod_{i=1}^{g} v_{i},
\eeq
and therefore the argument of the Poincar\'e series is given by
\beq
\frac{det(\Im(\tau))}{det(\Im(\tau_{22}))} = \prod_{i=1}^{r} v_{i}.
\eeq
This means that $\phi_{r}(\tau, s)$ is constructed by summing over the images under the modular group $\Gamma_{g}$
of a function $\varphi(v^{(r)}, s)$ of $U \backslash \mathbb{H}_{r}$, where $v^{(r)} = (v_{1},\dots ,v_{r})$.
\\
 If some of the Abelian coordinates  $v_{i} \rightarrow 0$, ($i = 1, \dots , r$), $\tau$ reaches the $(g - r)$-dimensional component of the boundary of $\mathbb{H}_{g}$,
which is stabilized by the parabolic subgroup $P_{g - r}$. The  left quotient
by $P_{g - r} \cap \Gamma_{g}$ in eq. (\ref{Poinc}) avoids   overcounting  modular transformations,
as $\tau$ reaches the $(g- r)$-dimensional component of the $\mathbb{H}_g$ boundary.

 Under suitable boundary conditions at the $r$-dimensional component of the $\Gamma_{g} \backslash \mathbb{H}_g$ boundary for  the modular invariant function $f(\tau)$, the
modular integral $I_{g,r}(s)$ inherits the $\phi_{r}(\tau,s)$ analytic properties in the complex variable $s$.
In Iwasawa coordinates, the modular integral acquires the following form
\beq
I_{g,r}(s) = \int_{ \Gamma_{g} \backslash \mathbb{H}_{g}} d\vec{w} \, d\vec{u} \, \prod_{i=1}^{g} dv_{i} \,  v_{i}^{i - g - 2} f(\tau) \phi_{r}(\tau,s),
\label{igrs}
\eeq
where the Jacobian determinant $J$ of the transformation (\ref{Iwcoord}) is given by
\beq
J = \prod_{i=1}^{g}v_{i}^{i - 1}.
\eeq
By using the unfolding trick, eq. (\ref{igrs}) reduces to
\bea
I_{g,r}(s) &=& \int_{\left( P_{g - r} \cap  \Gamma_{g} \right) \backslash \mathbb{H}_{g}}  d\vec{w} \, d\vec{u} \prod_{i=1}^{g} \, dv_{i} \,
\, v_{i}^{i - g - 2} f(\vec{v},\vec{w},\vec{u}) \, \varphi_{r}(v^{(r)},s) \nn \\
&=& \int_{0}^{\infty}dv_{1} \dots \ \int_{0}^{\infty}dv_{g} \prod_{i=1}^{g} \, v_{i}^{i - g - 2}  \varphi_{r}(v^{(r)},s) \int_{ \left( P_{g - r}
\cap  \Gamma_{g} \right) \backslash  U}  d\vec{w} \, d\vec{u} f(\vec{v},\vec{w},\vec{u}) \nn \\
&=& \int_{0}^{\infty}dv_{1} \dots \ \int_{0}^{\infty}dv_{g} \prod_{i=1}^{g} \, v_{i}^{i - g - 2}  \varphi_{r}(v^{(r)},s) < f >_{ \left( P_{g - r}
\cap  \Gamma_{g} \right) \backslash  U }(\vec{v}) \label{unf}
\eea
$< f >_{ \left( P_{r} \cap  \Gamma_{g} \right) \backslash  U }(\vec{v})$ is the $f$ average along the unipotent flow computed with the
$\Gamma_{g}$-invariant metric
\beq
< f >_{ \left( P_{g - r} \cap  \Gamma_{g} \right) \backslash  U }(\vec{v}) = :  \int_{ \left( P_{g - r} \cap  \Gamma_{g} \right) \backslash  U}  d\vec{w} \, d\vec{u} f(\vec{v},\vec{w},\vec{u}).
\eeq
It is interesting to check whether the  meromorphic structure  of $I_{g,r}(s)$ constrains (some) $v_{i} \rightarrow 0$ limits of the
unipotent average of the modular invariant $f$.

We start by studying the  modular integral of the product of a $Sp(2g,\mathbb{Z})$ invariant function $f(\tau)$
 times the  $E_{g, r}(\tau,s)$ Eisenstein series
\begin{eqnarray}
&& I_{g,r}(s) =  \int_{\Gamma_{g} \backslash \mathbb{H}_{g}} d\mu f(\tau) E_{r}(\tau,s), \label{modularIE}\\
&& E_{g, r}(\tau,s) =  \sum_{\Gamma_{g} \cap P_{g - r}  \backslash \Gamma_{g}} \left( \frac{det(\Im(\gamma(\tau)))}{det(\Im((\gamma (\tau)_{22}))}  \right)^{s}.
\end{eqnarray}
Analytic properties of the Eisentein series $E_{g, r}(\tau,s)$
are given for example in  \cite{Yamazaki} (theorem 2.2 and theorem 2.3).
Following \cite{Yamazaki}, the genus $g$ dressed Eisenstein  series  $\mathcal{E}_{g, r}(\tau,s)$, $r = 2,\dots ,g - 1$ 
\beq
\mathcal{E}_{g, r}(\tau,s) = \prod_{i=1}^{r}\zeta^{*}(2s + 1 -i) \prod_{i=1}^{[r/2]} \zeta^{*}(4s - 2g + 2r - 2i)E_{g, r}(\tau,s) \label{dress}
\eeq
is meromorphic in the complex variable $s$, with a simple pole in $s = g  - (r - 1)/2$
with residue
$$\frac{1}{2}\prod_{j=2}^{r} \zeta^{*}(j)  \prod_{j=1}^{[r/2]} \zeta^{*}(2g -  2r + 2j + 1 ),$$
where $\zeta^{*}(s) = \pi^{-s/2} \Gamma(s/2)\zeta(s)$, and $[x]$ denotes the integral part of the real number $x$.\\
Analytic properties of $\mathcal{E}_{g, r}(\tau,s)$ and eq. (\ref{dress})
imply that $E_{g, r}(\tau,s)$ is meromorphic on the $s$ plane with a simple pole
in $s = g  - (r - 1)/2$ with residue
\beq Res_{s \rightarrow g  - (r - 1)/2} E_{g, r}(\tau,s) =
\frac{\frac{1}{2}\prod_{j=2}^{r} \zeta^{*}(j)  \prod_{j=1}^{[r/2]} \zeta^{*}(2g -  2r + 2j + 1 )}{\prod_{i=1}^{r} \zeta^{*}(2g - r + 2 -i)
\prod_{i=1}^{[r/2]} \zeta^{*}(2g + 2 - 2i)}. \label{Eresidue}
\eeq
Moreover, $E_{g, r}(\tau,s)$ has poles in $s = \frac{\rho}{2} + \frac{i-1}{2}$ and $s = \frac{\rho}{4} + \frac{g - r + i}{2}$,
for $i = 1, \dots, r$, where   $\rho$ are  the non trivial zeros of the Riemann zeta function, $\zeta^{*}(\rho) = 0$.\\
In the $r = 1$ case, $E_{g, 1}(\tau,s)$ has a simple pole in $s = g$ with residue $1/\zeta^{*}(2g)$ and poles in $\rho/2$,
where $\rho$'s are the non trivial zeros of the Riemann zeta function, (\cite{Yamazaki}, theorem 2.3).
After the unfolding trick illustrated in (\ref{unf}), the modular integral (\ref{modularIE})
reduces to
\beq
I_{g,r}(s) = \int_{ (\Gamma_{g} \cap P_{g -r}) \backslash \mathbb{H}_g}  \prod_{i=1}^{r} dv_{i} \, v_{i}^{i - g - 2 + s}
\prod_{j= r + 1}^{g} dv_{j} \, v_{j}^{j - g - 2} d\vec{u} \, d\vec{w}  f (\vec{v}, \vec{u}, \vec{w} ). \label{Unfolded}
\eeq

\subsection{Modular holography, equidistribution  convergence rates and  the Riemann hypothesis}
In this section we develop a method for reducing a given modular integral of a  $Sp(2g,\mathbb{Z})$-invariant function $f(\tau)$  over a fundamental
domain $\Gamma_g \backslash  \mathbb{H}_g$ into an integral of the $f(\tau)$ average along some unipotent directions,  over the $(g - 1)$-dimensional
component $F_{g-1}$ of the boundary of $\mathbb{H}_g$.
By iterating this method one can then reduce the original genus $g$ integral into the limit towards the zero-dimensional component
$F_0$ of the $\mathbb{H}_g$ boundary of the average of $f(\tau)$ along all the unipotent directions in $\mathbb{H}_g$. Moreover, we will prove that
 equidistribution convergence rates for the various lower dimensional integrals we obtain in the process are related to
 the Riemann hypothesis.

\

The $(g -1)$-component $F_{g-1}$ of the   $\Gamma_g \backslash \mathbb{H}_g$  boundary is given by 
\beq
F_{g-1}  =
\begin{pmatrix}
i \infty & 0 \\
   0    &  \tau_{g - 1}
\end{pmatrix},  \qquad \tau_{g - 1} \in \mathbb{H}_{g - 1}.  \label{F}
\eeq
In Iwasawa coordinates $\mathbb{H}_g$ is given by
\beq
\tau = W + i UV^{2}U^{t},
\eeq
and with the convention for  $U$  to be a  upper triangular matrix,
one finds for the first $\tau$ entry
\beq
\tau_{11} = w_{11} + i(v_{1} +\bar u  V^2_{g-1} \bar u^t),
\eeq
where  $\bar u$ is the $(g-1)$-dimensional row vector with components $\bar{u} = (u_{12}, \dots, u_{1g})$  and $V^2_{g-1}=\rm{diag} \{v_2,\ldots, v_g\}$.

For $r =1$, the  unfolding equation  (\ref{Unfolded}) gives
\beq
\int_{ \Gamma_g \backslash \mathbb{H}_g} d\mu \, f(\tau) E_{g, 1}(\tau, s)  = \int_{ (\Gamma_{g} \cap P_{g - 1}) \backslash \mathbb{H}_g}  dv_{1} \, v_{1}^{s  - g - 1} \prod_{j= 2}^{g} dv_{j} \, v_{j}^{j - g - 2} d\vec{u} \, d\vec{w}  f (\vec{v}, \vec{u}, \vec{w} ), \label{unff}
\eeq
where the $P_{g - 1} \cap \Gamma_g $ parabolic subgroup which stabilizes the rational component $F_{g -1}$ of the $\mathbb{H}_g$ boundary
is given by the following matrices
\beq
\begin{pmatrix}
 1 &   *  &  * & * \\
 * & a &  0 &  b \\
 0 &  0  &  1 & *\\
 * & c &  0 &  d
 \end{pmatrix},
\qquad
\begin{pmatrix}
a & b \\
c & d
\end{pmatrix}
 \in \Gamma_{g -1}.
\eeq
For the  generic matrix in $P_{g -1}$ one has the decomposition
\beq\label{matrix}
\begin{pmatrix}
 1 &  m &   q  &  n  \\
 0 &  a &  n^{t}  &  b \\
 0 &   0 &  1  &  0\\
 0 &  c &  - m^{t}  &  d
\end{pmatrix}
= g_1 \cdot g_2 \cdot g_3,
\eeq
with, (see for example \cite{Huleketall}),
\beq
g_1 =
\begin{pmatrix}
 1 & 0 & 0 & 0 \\
 0 &  a & 0 &  b \\
 0 & 0 & 1 & 0 \\
 0 & c & 0 & d
\end{pmatrix} ,
\qquad
\begin{pmatrix}
a & b \\
c & d
\end{pmatrix}
 \in \Gamma_{g -1},
\eeq
\beq
g_2 =
\begin{pmatrix}
 1 & m & 0 & n \\
 0 & 1 & n^t & 0 \\
 0 & 0 & 1 & 0\\
 0 & 0 & - m^t & 1
\end{pmatrix}
\qquad
m,n \in Mat(1\times (g - 1), \mathbb{Z}),
\eeq
and
\beq
g_3 =
\begin{pmatrix}
 1 & 0 & q & 0 \\
 0 & 1 & 0 & 0 \\
 0 & 0 & 1 & 0 \\
 0 & 0 & 0 & 1
\end{pmatrix}
\qquad
q \in \mathbb{Z}.
\eeq
On $\tau \in \mathbb{H}_g$
\beq
\tau =
\begin{pmatrix}
\tau_1 & \tau_2  \\
\tau_{2}^{t} & \tau_3
\end{pmatrix}
\qquad  \tau_1 \in \mathbb{H}_{1}, \, \tau_2 \in Mat(1\times (g-1), \mathbb{C}), \,  \tau_3 \in \mathbb{H}_{g - 1},
\eeq
the action of $g_1$, $g_2$ and $g_3$ is given by  \cite{Huleketall}
\begin{eqnarray}
&& g_1(\tau) =
\begin{pmatrix}
\tau_1 -  \tau_{2}(c\tau_3 + d)^{-1}c\tau_2^t &  *  \\
(c\tau_3 + d)^{-1} \tau_{2}^t  & (a\tau_3 + b)(c\tau_3 + d)^{-1}
\end{pmatrix},\\
&& g_2(\tau) =
\begin{pmatrix}
 \tau_1^{'} &  *  \\
 \tau_{2}^{t} + m^t \tau_1 + n^t   &   \tau_{3}
\end{pmatrix}, \qquad   \tau_{1}^{'} = \tau_{1}  + m\tau_{3}m^{t} + m^t \tau_{2} + (m^t\tau_{2})^{t} + n m^{t},\\
&& g_3(\tau) =
\begin{pmatrix}
 \tau_1+q &  *  \\
 \tau_{2}^{t}    &   \tau_{3}
\end{pmatrix},
\end{eqnarray}
where the entries $*$ are given by symmetry of $\tau$.

The above decomposition shows that $P_{g -1} \cap \Gamma_{g}$ contains $\Gamma_{g -1}$ as a subgroup acting on $\tau_{3} \in \mathbb{H}_g$.
Therefore, one in eq. (\ref{unff}) can take the average of $f(\tau)$
along the appropriate unipotent directions, in order to obtain a $\Gamma_{g-1}$-invariant  function $< f >_{2g - 1 }(v_{1}, \tau_{3})$,
with $\tau_3 \in \mathbb{H}_{g - 1}$.
This can be obtained as follows. Specializing to the case $r=1$, it is convenient to express the Iwasawa parametrization of
$\tau$ by evidencing the $(g-1)$-structures:
\begin{eqnarray}
U=: \begin{pmatrix}
1 &  \bar u \\
\bar{0}^t & U_{g-1}
\end{pmatrix}, \qquad
V^2={\rm diag}\{v_1; \bar v\},
\end{eqnarray}
where  $\bar u$ is the $(g-1)$-dimensional row vector $\bar{u} = (u_{12},\dots , u_{1g})$,  and $\bar v = (v_2,\ldots, v_g) $.
Then (\ref{Iwcoord}) takes the form
\begin{eqnarray}
\tau=\tau_{g-1} +\begin{pmatrix} w_{11} & \bar w \\ \bar w^t & \mathbb{O} \end{pmatrix}
+i \begin{pmatrix} v_1+\bar u  V^2_{g-1} \bar u^t & \quad\ \bar u  V^2_{g-1} U^t_{g-1} \\ U_{g-1} V^2_{g-1} \bar u^t & \mathbb{O}  \end{pmatrix},
\end{eqnarray}
where
\begin{eqnarray}
\tau_{g-1}=\begin{pmatrix} 0 & \bar 0 \\ \bar 0^t & \tau_3 \end{pmatrix}, 
\end{eqnarray}
$\bar{w} = (w_{12},\dots , w_{1g})$     $U_{g-1}$ is the minor of $U_{11}$, and   $\mathbb{O}$ is the null squared
$(g - 1)$-dimensional  matrix.
Using this in eq. (\ref{unff}) one gets
\beq
\int_{ \Gamma_g \backslash \mathbb{H}_g} d\mu_{g} \, f(\tau) E_{g, 1}(\tau, s)  = \int_{0}^{\infty} d v_{1} v_{1}^{s  - g - 1}
\int_{ \Gamma_{g -1} \backslash \mathbb{H}_{g - 1} }d\mu_{g - 1}  < f >_{2g - 1}(v_1 ,  \tau_3  ), \label{unf3}
\eeq
where in the average $< f >_{2g - 1}(v_1 ,  \tau_3  )$ the integration over the $(2g - 1)$ unipotent coordinates $w_{11}, \bar u, \bar w$ that
reduce $\mathbb{H}_g \rightarrow \mathbb{H}_{g -1}$ takes into account the identifications induced by the left quotient by the parabolic
subgroup $P_{g -1}$.\\
In the r.h.s. of eq. (\ref{unf3}) the Mellin transform of the following function appears
\beq
\mathcal{F}_{g-1}(v_1) = : \frac{1}{v_{1}^{g}} \int_{ \Gamma_{g -1}
\backslash \mathbb{H}_{g - 1} }d\mu_{g - 1}  < f >_{2g - 1}(v_1 ,  \tau_3  ).
\eeq
The following condition
\beq
\lim_{v_1 \rightarrow 0}  \int_{ \Gamma_{g -1} \backslash \mathbb{H}_{g - 1} }d\mu_{g - 1}
< f >_{2g - 1}(v_1 ,  \tau_3  ) = C_{g},
\eeq
reproduces the simple pole in $s = g$ of the genus  $g$, $r =1$ Eisenstein series $E_{g, 1}(\tau, s)$,
whose analytic properties are given before the end of the previous section. One has
\beq
Res_{s \rightarrow g}E_{g,1}(\tau , s) = \frac{1}{ 2\zeta^{*}(2g)} = \frac{Vol(\Gamma_{g - 1} \backslash \mathbb{H}_{g-1})}{2 Vol(\Gamma_{g}
\backslash \mathbb{H}_{g})},
\eeq
since
\beq
Vol(\Gamma_{g} \backslash \mathbb{H}_{g}) = 2 \prod_{k = 1}^{g} \zeta^{*}(2k).
\eeq
Therefore one has in eq. (\ref{unf3}) as $v_1 \rightarrow 0$
\beq
 \int_{ \Gamma_{g -1} \backslash \mathbb{H}_{g - 1} }d\mu_{g - 1}
 < f >_{2g - 1}(v_1 ,  \tau_3  ) \sim \frac{ Vol(\Gamma_{g - 1} \backslash \mathbb{H}_{g-1})}{ Vol(\Gamma_{g} \backslash \mathbb{H}_{g}) }
\int_{ \Gamma_g \backslash \mathbb{H}_g} d\mu_{g} \, f(\tau)  \qquad v_1 \rightarrow 0 \label{lim}
\eeq

We can also obtain the $v_{1} \rightarrow 0$ convergence rate in (\ref{lim}), which is quite interestingly  related to the Riemann hypothesis.
Due to the location of the poles of
$E_{g,1}(\tau, s)$, the Mellin transform in the
r.h.s. of eq. (\ref{unf3}) is analytic on the half-plane $\Re(s) > \frac{\Theta}{2}$,
except for a simple pole in $s = g$.
  $\Theta = Sup\{\Re(\rho) | \zeta^{*}(\rho) = 0 \}$ is the superior of the real part of the non trivial zeros of the Riemann
zeta function. By inverse Mallin transform argument, one then finds the following $v_1 \rightarrow 0$ error estimate 
\beq
  \int_{ \Gamma_{g -1} \backslash \mathbb{H}_{g - 1} }d\mu_{g - 1}
< f >_{2g - 1}(v_1 ,  \tau_3  ) \sim 
  \frac{ Vol(\Gamma_{g - 1} \backslash \mathbb{H}_{g-1}) }{ Vol(\Gamma_{g} \backslash \mathbb{H}_{g}) }
\int_{ \Gamma_g \backslash \mathbb{H}_g} d\mu_{g} \, f(\tau) + O(v_{1}^{{g} - \frac{\Theta}{2}} ). \label{lim-rate}
\eeq
Notice that the above convergence rate is $O(v_{1}^{{g} - \frac{1}{4}} )$ iff the Riemann hypothesis is true!
This result for the $Sp(2g,\mathbb{Z})$  equidistribution rate corresponds
to the  $3/4$ rate condition \cite{ZagierE}  in the $Sl(2,\mathbb{Z})$  case for functions of rapid decay   $z \rightarrow i\infty$, which is indeed
recovered in (\ref{lim-rate}) by setting $g=1$.

\

By iteration of the above method it looks that one recovers equidistribution of the $f(\tau)$ average over the  full set of  unipotent
coordinates of $\mathbb{H}_g$ in the zero-dimensional boundary limit. Furthermore, one obtains the convergence rates for the $\mathbb{H}_g$ abelian coordinates, and an intriguing  connection among the values of the powers of the error terms and the Riemann hypothesis.
A computation of those values which do not beg on the Riemann zeta function would prove or disprove the
Riemann hypothesis. It would be therefore quite interesting to be able to map the problem in
String Theory terms and gain a new angle from which to estimate those quantities.

\section{Conclusions and perspectives}
In this paper we have started to tackle the problem of determine equidistribution properties for  genus $g \ge 2$
$Sp(2g, \mathbb{Z})$ automorphic forms defined on the Siegel upper space $\mathbb{H}_g$. Our main aim is to
explore the potential applications to  genus $g \ge 2$ string amplitudes. For $g=1$ the relevance of uniform distribution
has been put  in evidence in \cite{Cardella:2008nz}, where it is shown how to determine  constraints
on the ultraviolet property of   non tachyonic closed string spectra \cite{Kutasov:1990sv}.

For genus $g=2$, the vacuum-to-vacuum superstring amplitudes has been computed in \cite{DP0}--\cite{DP5}, and recent proposals for higher
genus can be found, for example, in \cite{Cacciatori:2007vk}--\cite{Grushevsky:2008zm}. Then, it should be interesting to apply
the uniformization method in order to obtain constraints related to finiteness of genus $g$ closed string amplitudes. This is indeed
 under study at the moment. However, note that at the actual stage the uniformization methods can be applied to string theory only up to genus three.
Indeed, (super)string integrals must  be performed over the moduli space of  Riemann surfaces,
and for  genus higher than $g=3$ this task is  related to the Schottky problem! This suggests that it should be really interesting to try
to find equidistribution theorems valid for automorphic forms defined only over the Schottky locus.

Beyond physical applications, it is interesting to note the relations with number theory:  \cite{ZagierE} has shown that
the equidistribution rate is intimately related to the Riemann hypothesis, which is proven to hold true if one is able to show that
for a certain class of $Sl(2, \mathbb{Z})$ automorphic functions their horocycle convergence rate is $O(\Im (z)^{3/4})$. Here we have found that in the case of codimension 1
reductions, the same holds true for the $Sp(2g, \mathbb{Z})$ case with a convergence rate $O(\Im (z)^{(4g-1)/4})$. To our knowledge
this result is new.
This reduction
can be in principle reiterated to obtain higher codimension reductions. However, it is interesting to note that if one considers directly
$r$-codimensional reductions, with $r>1$, then new poles appears, which could provide new information. Thus, one should try to extend these
results to such general cases.

The physical and mathematical interests can be further interweaved by the following remark: There are known examples where the
convergence rate corresponds to physical quantities, see \cite{ACER}, for example. This suggests to investigate such correspondence
more deeply in order to relate equidistribution rates to string quantities which, after dualities, could be put in a calculable
form. This could provide a way to gain a  string theory  perspective on  the Riemann hypothesis.

All these points are currently under investigation.

\section{Acknowledgments}
The authors would like to  thank Bert van Geemen for  enlightening discussions.
M.C. thanks Don Zagier for pointing out to him his result  on the connection between   horocycle average
 convergence rate and the Riemann hypothesis.  The work of M.C. is supported in part by the Italian MIUR-PRIN contract 20075ATT78 and in part
 by a  ESI visiting fellowship. M.C. thanks the ESI
 Schr\"odinger center for Mathematical Physics in Vienna for very nice  hospitality and support during the final stage of this work.

\appendix
\section{Appendix}

\subsection{ On $\mathbb{H}_g$ boundary components and their parabolic subgroups stabilizers}
In this section   we review the
structure of the boundary of $\mathbb{H}_g$. We will show that  $\partial \mathbb{H}_g$ is given by components of various dimensionality which  are stabilized by parabolic subgroups of $Sp(2g,\mathbb{R})$. The strategy in order to study the structure of $\partial \mathbb{H}_g$ \cite{Huleketall}   is to use the Cayley map to go from
$\mathbb{H}_g$  to the multidimensional generalization of the open unit Poincar\'e disc
$\mathbb{D}_g = \{z \in Mat(2g, \mathbb{C}) | z = z^{t},  z\bar{z} < \mathbb{I} \}$. One then takes the closure $\bar{\mathbb{D}}_g$,
and study the  boundary of $\Gamma_{g}  \backslash \bar{\mathbb{D}}_g$.
This is done by constructing an  equivariant map under $Sp(2g, \mathbb{R})$ which relates  boundary components
of $\Gamma_{g}  \backslash \bar{\mathbb{D}}_g$  to isotropic subspaces in $\mathbb{R}^{2g}$.
Since isotropic subspaces in $\mathbb{R}^{2g}$ are stabilized by parabolic subgroups of $Sp(2g, \mathbb{R})$, it follows that
boundary components of  $\Gamma_{g}  \backslash \bar{\mathbb{D}}_g$ are stabilized by parabolic elements
of  $Sp(2g, \mathbb{R})$. This constructions provides an explicit characterization of the
boundary components of $\Gamma_{g}  \backslash \bar{\mathbb{D}}_g$ and for each component  the explicit form
of the  parabolic subgroup stabilizer.

\

The Cayley map is defined as $z :  \mathbb{H}_g \rightarrow \mathbb{D}_g$
\beq
 z(\tau) = (\tau - i\mathbb{I})(\tau + i\mathbb{I})^{-1},
\eeq
where the bounded domain $\mathbb{D}_g = \{z \in Sym(2g, \mathbb{C}) | \, z\bar{z} < \mathbb{I} \}$ is the multidimensional analogous
of the Poincar\'e open disc. Let $\bar{\mathbb{D}}_g = \{z \in Sym(2g, \mathbb{C}) | \, z\bar{z} \le \mathbb{I} \}$ the closure of $\mathbb{D}_g$
and $\partial \bar{\mathbb{D}}_g = \bar{\mathbb{D}}_g - \mathbb{D}_g $ the boundary of $\mathbb{D}_g $.

We now define a map which  allows to explore the structure of the boundary of $ \Gamma_{g}  \backslash  \bar{\mathbb{D}}_g$.
It is defined as  $\Psi_{z}(w) :  \mathbb{C}^{g} \rightarrow \mathbb{C}^{g}$,
$ \Psi_{z}(w) = : i(wz + \bar{w})$, $z \in \bar{\mathbb{D}}_g$. $\Psi_{z}(w)$ enjoys the following property, let
$U(z) = : Ker( \Psi_{z}(w)) =   \{ w \in \mathbb{C}^g | \bar{w} = - wz \}$,
then $U(z) \ne 0$ iff $z \in \partial \bar{\mathbb{D}}_g$. Moreover, the image of $U(z)$ in $\mathbb{R}^{2g}$
through the map $\nu : \mathbb{C}^g \rightarrow \mathbb{R}^{2g}$,  $\nu_{i} = \frac{1}{2}\left( w_i + \bar{w}_i   \right)$,
$\nu_{g + i} = \frac{1}{2i}\left( w_i - \bar{w}_i   \right)$, $i = 1, \dots , g$,
is an isotropic space, (i.e. $< \nu, \tilde{\nu} > = \nu  J \tilde{\nu}^{t} = 0$, where $J$ is the standard symplectic form).

\

Let us start to show that $U(z) \ne 0$ iff $z \in \partial \bar{\mathbb{D}}_{g}$.
Suppose $U(z) \ne 0$, it means that there is a $z \in \bar{\mathbb{D}}_{g}$ such that $\bar{w} = - z w$ for some $w \in \mathbb{C}^{g}$.
Indeed, the above equation is satisfied by
\beq
z^{(k)} =   \begin{pmatrix}
    \mathbb{I}_{g - k} & 0 \\
    0      & z_{(k)}
\end{pmatrix}, \qquad  z_{(k)} \in \mathbb{D}_{k}
\eeq
and the $g$-dimensional   vector $w_{(k)} = (i\nu_{1}, \dots , i\nu_{g-k}, 0, \dots, 0)^{t}$, with pure imaginary entries
satisfies $\bar{w}_{(k)} = - z^{(k)}w_{(k)}$ for $k = 0, \dots, g$.
Notice that $w_{(k)}\in \mathbb{C}^{g}$ corresponds to the  real vector $\nu_{(k)} = (0, \dots, 0; \nu_{1}, \dots , \nu_{g-k}, 0,\ldots, 0)^{t} \in
\mathbb{R}^{2g}$, via the $\nu$ map, where the first $g$ entries are zero.\\
Moreover,
\beq
z^{(k)}\bar{z}^{(k)} - 1 =   \begin{pmatrix}
    0_{g - k} & 0 \\
    0   & z_{(k)}\bar{z}_{(k)}-\mathbb{I}_{k}
\end{pmatrix},
\eeq
is clearly a nonpositive matrix, therefore $z^{(k)} \in \partial \bar{\mathbb{D}}_{g}$. We see that $\partial \bar{\mathbb{D}}_{g}$
decomposes into $g$ components $(\partial \bar{\mathbb{D}}_{g})_{k} = \bar{\mathbb{D}}_{k}$, $k = 0, \dots, g - 1$.
Indeed, the boundaries related to the quotients of the Siegel spaces by  arithmetic subgroups  of $Sp(2g,\mathbb{Z})$
turn out to be rational component of $\partial (\bar{\mathbb{D}}_{g})_{k} = \bar{\mathbb{D}}_{k}$,
similarly to the $Sl(2,\mathbb{Z})$ case, where  cusps are given by rational points of the boundary of $\mathbb{H}_{1}$,
(see for example  \cite{Huleketall}  for definitions and characterizations of the rational component of
$\partial (\bar{\mathbb{D}}_{g})_{k} = \bar{\mathbb{D}}_{k}$).

\

Each boundary component $(\partial \bar{\mathbb{D}}_{g})_{k}$  is in correspondence to the real isotropic space in $\mathbb{R}^{2g}$
spanned by  the  vectors $\nu_{(k)} = (0, \dots, 0; \nu_{1}, \dots , \nu_{g-k}, 0,\ldots, 0)^{t}$,
through the map
\beq
\Psi_{z}(\nu) = 2i \nu
\begin{pmatrix}
  \mathbb{I} & - i\mathbb{I} \\
  \mathbb{I} &  i\mathbb{I}
\end{pmatrix}^{-1}
\begin{pmatrix}
    \mathbb{I} \\
    z
\end{pmatrix}.
\eeq
Since under
\beq
 g =
\begin{pmatrix}
   a &  b \\
    c     &  d
\end{pmatrix}
  \in Sp(2, \mathbb{R}),
 \eeq
one has
\beq
\begin{pmatrix}
    \mathbb{I} \\
    z
\end{pmatrix}
\rightarrow
 \begin{pmatrix}
   \mathbb{I} & - i\mathbb{I} \\
    \mathbb{I}     &  i\mathbb{I}
\end{pmatrix}
\begin{pmatrix}
   a &  b \\
    c     &  d
\end{pmatrix}
\begin{pmatrix}
   \mathbb{I} & - i \mathbb{I} \\
    \mathbb{I}     &  i\mathbb{I}
\end{pmatrix}^{-1}
\begin{pmatrix}
    \mathbb{I}\\
    z
\end{pmatrix},
\eeq
it follows that
\beq
\Psi_{g(z)}(\nu) = \Psi_{z}(\nu \cdot g). \label{equiv}
\eeq
Whenever $z \in \partial \bar{\mathbb{D}}_{g}$ is in the boundary of the closure of the $g$-dimensional disc,
$\nu \in \mathbb{R}^{2g}$ spans an isotropic subspace of $\mathbb{R}^{2g}$.
Since isotropic spaces are stabilized by parabolic elements of $Sp(2g,\mathbb{R})$ it follows from eq. (\ref{equiv}) that parabolic
subgroups stabilize the boundary. Moreover,  the boundary of $\bar{\mathbb{D}}_{g}$
decomposes into $g$ components $\partial \bar{\mathbb{D}}_{g} = \cup_{a = 0}^{g - 1} F_{a}$,
where
\beq
F_{a}  =   \begin{pmatrix}
    \mathbb{I}_{g - a} & 0 \\
    0      & z_{(a)}
\end{pmatrix}, \qquad  z_{(a)} \in \mathbb{D}_{a}.
\eeq
$F_{a}$, $a = 1, \dots, g -1$ are stabilized  by parabolic subgroups $Sp(2g, \mathbb{R})$ of corresponding dimensionality.

\subsection{$g = 1,2$ examples}
Let us start by applying the  results of the previous section to the genus $g = 1$ case in order to recover the parabolic
subgroup $\Gamma_{\infty} \subset Sl(2,\mathbb{Z})$ cusp stabilizer.
Let $z \in \bar{\mathbb{D}}_1$, there is only one boundary component $z = 1$ of dimension zero, (which is the image through the Cayley map of
the $\tau = i\infty$
cusp in $\mathbb{H}_1$). Thus $r = 0$, and the isotropic subspace in $\mathbb{R}^2$ related to the $z = 1$ boundary  is given by  vectors of
the form $(0,\nu)$.
The $Sl(2,\mathbb{R})$ equivariant map is given by
\beq
 \Psi_{z}(\nu) = 2i \nu
  \begin{pmatrix}
   1 & - i \\
    1     &  i
\end{pmatrix}^{-1}
\begin{pmatrix}
    1 \\
    z
\end{pmatrix},
\eeq
which under $g \in Sl(2,\mathbb{R})$,  $ \Psi_{z}(\nu \cdot g) = \Psi_{g(z)}(\nu)$.
Therefore  a $Sl(2,\mathbb{Z})$ modular transformation $\gamma$
\beq
\gamma =
 \begin{pmatrix}
   a &  b \\
   c     &  d
\end{pmatrix}
\eeq
stabilizes the $z = 1 $
cusp iff
\beq
(0,\nu)
 \begin{pmatrix}
   a &  b \\
    c     &  d
\end{pmatrix}
= (\nu c, \nu d )
= (0, \nu).
\eeq
This leads to $c= 0$, $d =1$, $a = 1$, $b \in \mathbb{Z}$,
and one recovers the well known $g =1$ parabolic\footnote{A matrix $\gamma \in Sl(2,\mathbb{Z})$ is parabolic iff $Tr(\gamma ) = 2$.}
subgroup $\Gamma_{\infty} \subset \Gamma$
of the modular group $\Gamma \sim Sl(2, \mathbb{Z})$, given by the unipotent matrices
\beq
 \begin{pmatrix}
   1 &  b \\
    0     &  1
\end{pmatrix}, \qquad  b \in \mathbb{Z}.
\eeq

\

We consider now the $g =2$ case. $\bar{\mathbb{D}}_2$ boundary has two components,
the zero dimensional component  $F_{0}$
\beq
z =
 \begin{pmatrix}
   1 &  0 \\
    0     &  1
\end{pmatrix},
\eeq
and the one dimensional component  $F_{1}$
\beq
z  =
  \begin{pmatrix}
   1    & 0 \\
   0    &  z
\end{pmatrix}, \qquad  z \in \bar{\mathbb{D}}_{1}.
\eeq
Let us start by obtaining the parabolic subgroup of $Sp(4,\mathbb{Z})$
which stabilizes  $F_{0}$. According to what explained in the previous section,
the isotropic space in $\mathbb{R}^4$ related to $F_0$
is given by vectors of the form, (r = 0)
\beq
\nu = (0,0, \nu_1 , \nu_2 ).
\eeq
The parabolic subgroup $P_0$  is obtained by the following condition
\beq
(0,0, \nu_1 , \nu_2 )
= (0,0, \nu_1 , \nu_2 )
 \begin{pmatrix}
   a_1 &  a_2 &  b_1  &  b_2 \\
 a_4 &  a_3 &  b_4  &  b_3  \\
c_1 &  c_2 &  d_1  &  d_2 \\
c_4 &  c_3 &  d_4  &  d_3
\end{pmatrix},
\eeq
which implies that $P_0$ is given by  matrices in $Sp(4,\mathbb{Z})$ of the form
\beq
 \begin{pmatrix}
    \mathbb{I}    &  b \\
    0      &  \mathbb{I}
\end{pmatrix},
\eeq
where $b \in Mat(2,\mathbb{R})$ symmetric $b = b^t$. \\
In a similar way, the isotropic space related to $F_1$ is
\beq
\nu = (0,0, \nu_1 , 0).
\eeq
The parabolic group $P_1$ is thus determined by the condition
\beq
(0,0, \nu_1 , 0 )
= (0,0, \nu_1 , 0 )
 \begin{pmatrix}
   a_1 &  a_2 &  b_1  &  b_2 \\
 a_4 &  a_3 &  b_4  &  b_3  \\
c_1 &  c_2 &  d_1  &  d_2 \\
c_4 &  c_3 &  d_4  &  d_3
\end{pmatrix},
\eeq
which gives $c_1=c_2=d_2=0$, $d_1=1$. Imposing the $Sp(4,\mathbb{Z})$ conditions gives (\ref{matrix}).

\end{document}